\documentclass[aps,twocolumn,showpacs,showkeys]{revtex4}
\usepackage{psfig,amssymb,graphicx}
\begin{document}
%%%%%%%%%%%%%%%%%%%%%%%%%%%%%%%%%%%%%%%%%%%%%%%%%%%%%%%%%%%%%
\title{Radiation to atom quantum mapping by collective recoil
in Bose-Einstein condensate} \date{\today}
\author{Matteo G. A. Paris} \email{paris@unipv.it} \homepage{www.qubit.it/~paris}
\affiliation{Quantum Optics \& Information Group, INFM Unit\`a di Pavia, Italia}
\author{Mary Cola, Nicola Piovella, Rodolfo Bonifacio}
\affiliation{Dipartimento di Fisica and
Unit\`a INFM, Universit\'a di Milano, Italia.}
%%%%%%%%%%%%%%%%%%%%%%%%%%%%%%%%%%%%%%%%%%%%%%%%%%%%%%%%%%%%%
\begin{abstract}
We propose an experiment to realize {\em radiation to atom}
continuous variable quantum mapping, {\em i.e.} to teleport the
quantum state of a single mode radiation field onto the collective
state of atoms with a given momentum out of a Bose-Einstein
condensate. The atoms-radiation entanglement needed for the
teleportation protocol is established through the interaction of a
single mode with the condensate in presence of a strong far
off-resonant pump laser, whereas the coherent atomic displacement
is obtained by the same interaction with the radiation in a
classical coherent field.  In principle, verification of the
protocol requires a joint measurement on the recoiling atoms and
the condensate, however, a partial verification involving
populations, {\em i.e.} diagonal matrix elements may be obtained
through counting atoms experiments.
\end{abstract}
%%%%%%%%%%%%%%%%%%%%%%%%%%%%%%%%%%%%%%%%%%%%%%%%%%%%%%%%%%%%%
 \maketitle
%%%%%%%%%%%%%%%%%%%%%%%%%%%%%%%%%%%%%%%%%%%%%%%%%%%%%%%%%%%%%
Entanglement is a crucial resource in the manipulation of quantum
information, and quantum teleportation \cite{tel,Braun1} is
perhaps the most impressive example of quantum protocol based on
entanglement. Teleportation is the transferral of (quantum)
information between two distant parties that share entanglement.
There is no physical move of the system from one player to the
other, and indeed the two parties need not even know each other's
locations. Only classical information is actually exchanged
between the parties. However, due to entanglement, the quantum
state of the system at the transmitter location (say Alice) is
mapped onto a different physical system at the receiver location
(say Bob). The information transferral is {\em blind}, {\em i.e.}
the protocol should work also when the state to be teleported is
completely unknown to both the sender and the receiver. Several
teleportation protocols have been suggested either for qubits and
continuous variable systems
\cite{zei,dem,nmr,shi,sci,gis,kim,bab,bow}, including interspecies
teleportation of atomic spin onto polarization states of light
\cite{pol}.
\par
\begin{figure}[h!]
\centerline{\psfig{file=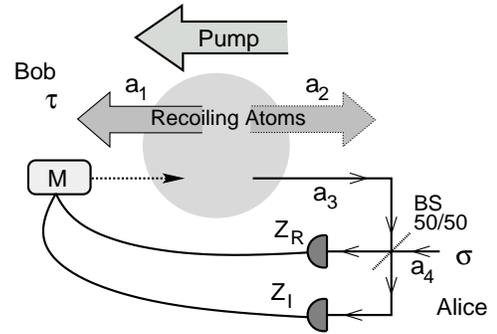,width=65mm}} 
\caption{ Schematic
diagram of the proposed experiment to realize {\em radiation to
atom} continuous variable quantum mapping, {\em i.e.}
teleportation of the quantum state of a single mode radiation
field onto the collective state of atoms with a given momentum out
of a Bose-Einstein condensate.  The experiment proceeds as
follows: the atomic mode $a_1$ and the radiation mode $a_3$ are
entangled through the interaction of the light mode with the
condensate in presence of a strong far off-resonant pump laser
(CARL dynamics).  The outgoing radiation mode $a_3$ is then mixed
(in a balanced beam splitter) at the sender' location (Alice) with
another radiation mode $a_4$, excited in the state $\sigma$, which
we want to teleport, and the joint measurement of a couple of
two-mode quadratures is performed. The result of the measurement
is sent to the receiver's location (Bob), where the corresponding
coherent atomic displacement is performed. The latter is obtained
through the same CARL interaction, by injecting a suitably
modulated coherent pulse (M denotes a modulator). The overall
dynamics is such that the ensemble of recoiling atoms in the mode
$a_1$ is described by the density matrix $\tau$, which approaches
$\sigma$ in the limit of high entanglement {\em i.e.} high gain of
the CARL interactions. \label{f:schema}}
\end{figure}

In this letter we propose a novel scheme to realize radiation to
atom quantum state mapping {\em i.e.} the interspecies
teleportation of the quantum state of a single mode radiation
field onto the the collective state of atoms with a given momentum
out of a Bose-Einstein condensate. The four basic ingredients of a
quantum teleportation experiment are the following: i) an
entangled state shared between two parties; ii) a joint {\em Bell
measurement} performed on the system whose state is to be
teleported and on one subsystems of the entangled state; iii) a
device able to perform a given class of unitary transformation,
{\em conditioned} to the results of the joint measurement; iv) a
readout system to verify teleportation. In the following we
describe the above points for our teleportation protocol, and
discuss the feasibility conditions of our proposal. The setup is
schematically illustrated in Fig. \ref{f:schema}.
%%%%%%%%%%%%%%%%%%%%%%%%%%%%%%%%%%%%%%%%%%%%%%%%%%%%%%%%%%%%%
%%%%%%%%%%%%%%%%%%%%%%%%%%%%%%%%%%%%%%%%%%%%%%%%%%%%%%%%%%%%%
\par
The entangled state supporting teleportation will be a {\em twin-beam-like}
state of a radiation mode and a collective mode of atoms with a given momentum
out of a Bose-Einstein condensate. It is given by the quantum CARL model for
interaction of a Bose Einstein condensate (BEC) with a single-mode quantized
radiation field in the presence of a strong far off-resonant pump laser. The
starting point of such a model is the classical Hamiltonian for $N$ two-level
atoms exposed to an off-resonant pump laser, whose electric field $\vec E_0=
\hat e{\cal E}_0\cos(\vec k_2\cdot\vec x-\omega_2t)$ is polarized along $\hat
e$, propagates along the direction of $\vec k_2$ and has a frequency $\omega_2
=ck_2$ with a detuning from the atomic resonance, $\Delta_{20}=\omega_2-
\omega_0$, much larger than the natural linewidth of the atomic transition,
$\gamma$. The atoms scatter a single-mode field circulating in a high-$Q$ ring-cavity,
with frequency $\omega_1$, wavenumber $\vec k_1$ making an angle $\phi$ with $\vec
k_2$ and electric field $\vec E=(\hat e/2)[{\cal E}(t)e^{i(\vec k_1\cdot\vec
x-\omega_1 t)}+{\rm c.c.}]$ with the same polarization of the pump field.
By adiabatically eliminating the internal atomic degree of freedom, the following
CARL Hamiltonian can be derived \cite{car}
$$H=\sum_{j=1}^N\left[\omega_r p_j^2-ig\left(a e^{i\theta_j}-{\rm
c.c.}\right)\right] -\Delta |a|^2\:,$$
where $\omega_r=\hbar q^2/2M$ is the recoil frequency, $M$ is the
atomic mass, $q=|\vec q|$ and $\vec q=\vec k_1-\vec k_2$ is the
difference between the scattered and the incident wave vectors,
$\theta_j=qz_j$ and $p_j=p_{zj}/\hbar q$ are the dimensionless
position and momentum of the $j$-th atom along the axis $z$
directed along $\vec q$, $g=(\Omega_0/2\Delta_{20})(\omega_2
d^2/2\hbar\epsilon_0 V)^{1/2}$, $a=-i(\epsilon_0 V/2\hbar
\omega_2)^{1/2}{\cal E}e^{i\Delta t}$, $\Delta=\omega_2-
\omega_1$, $\Omega_0=d{\cal E}_0/\hbar$ is the Rabi frequency of
the pump, $V$ is the interaction volume, $d$ is the atomic dipole
and $\epsilon_0$ is the permittivity of the free space.
\par
In order to quantize both the radiation field and the
center-of-mass motion of the atoms,  $\theta_j$, $p_j$ and $a$ are
considered as quantum operators satisfying the canonical
commutation relations $[\theta_j,p_{j'}]=i\delta_{j,j'}$ and
$[a,a^{\dag}]=1$. The model is then second quantized introducing
the atomic field operator $\Psi(\theta)$ with equal-time
commutation relations
$[\Psi(\theta),\Psi^{\dag}(\theta')]=\delta(\theta-\theta')$,
$[\Psi(\theta),\Psi(\theta')]=0$ and the normalization
condition
$\int_0^{2\pi}d\theta\Psi(\theta)^{\dag}\Psi(\theta)=N$.
Creation and annihilation operators are introduced for atoms with
a definite momentum $p$, i.e. $\Psi(\theta)=\sum_m c_m
u_m(\theta)$, where $u_m(\theta)=\exp(im\theta)/\sqrt{2\pi}$ and
$c_m$ are bosonic operators obeying the commutation relations
$[c_m,c^{\dag}_n]=\delta_{m,n}$ and $[c_m,c_n]=0$, and
$c_m^{\dag}$ creates an atom with momentum $p=m$ in $\hbar q$
unit. This description of the atomic motion in a BEC assumes that
the atoms are delocalized inside the condensate and that, at zero
temperature, the momentum uncertainty
$\sigma_{p_z}\approx\hbar/\sigma_z$ can be neglected with respect
to $\hbar q$. The approximation is valid for $L\gg\lambda$, if
$\sigma_z\approx L$, where $L$ is the size of the condensate. The
Hamiltonian for the second quantized model becomes \cite{las}
\begin{equation}
H= \sum_{n=-\infty}^\infty\left\{ \omega_r n^2
c_n^{\dag}c_n+ig\left(a^{\dag}c_{n}^{\dag}c_{n+1}-{\rm
h.c.}\right)\right\} -\Delta a^{\dag}a \label{ham2}
\end{equation}
Notice that the Hamiltonian (\ref{ham2}) commutes with
the number of atoms, $N=\sum_n c_n^{\dag}c_n$, and the total
momentum, $P=a^{\dag}a+\sum_n nc_n^{\dag}c_n$. Let us now consider
the equilibrium state with no field and all the atoms at
rest, i.e. in momentum state with $n=0$. In the linear regime, we
neglect the atomic depletion of the ground state $n=0$, taking
$c_0\approx\sqrt{N}$ as a $c$-number, and we consider only the
transitions induced by the radiation field from the state $n=0$
toward the levels $n=-1$ and $n=1$. Introducing the operators
$a_1=c_{-1}\exp(i\Delta t)$, $a_2=c_{1}\exp(-i\Delta t)$ and
$a_3=a\exp(-i\Delta t)$, the atomic field operator reduces to
\begin{equation}
\Psi(\theta,t)\approx \frac{1} {\sqrt{2\pi}} \left\{ \sqrt
N+a_1(t) e^{-i(\theta+\Delta t)}+a_2(t) e^{i(\theta+\Delta
t)}\right\} \label{psi:lin}
\end{equation}
and the Hamiltonian (\ref{ham2}) reduces to the effective
Hamiltonian \cite{moo}
\begin{equation}
H_0=\delta_+ a_2^{\dag}a_2 -\delta_- a_1^{\dag}a_1
+ig\sqrt{N}[(a_1^{\dag}+a_2)a_3^{\dag}-{\rm h.c.}],\label{ham3}
\end{equation}
where $\delta_{\pm}=\Delta\pm \omega_r$. Hence, the dynamics of
the system is that of three parametrically coupled harmonic
oscillators $a_1$, $a_2$ and $a_3$, which obey the
commutation rules $[a_i,a_j]=0$ and
$[a_i,a_j^{\dag}]=\delta_{i,j}$ for $i,j=1,2,3$. Notice that the
Hamiltonian (\ref{ham3}) admits $C=a_2^{\dag}a_2-a_1^{\dag} a_1+
a_3^{\dag}a_3$ as a constant of motion.
\par
Hence, in the linear regime, the quantum CARL Hamiltonian reduces
to that for three coupled modes, the first two modes $a_1$ and
$a_2$ corresponding to atoms having lost or gained respectively a
two-photon recoil momentum $\hbar q$, and the third mode $a_3$
corresponding to the photons of the scattered field.
Starting from the vacuum $|{\bf 0}\rangle=
|0\rangle_1 |0\rangle_2 |0\rangle_3$ the state at a given time
is given by
\begin{equation}
|\Psi\rangle=\frac{1}{\sqrt{1+N_1}} \sum_{m,n=0}^\infty
\alpha^m\beta^n\: \sqrt{\frac{(m+n)!}{m!n!}}|m+n,m,n\rangle
\label{state:0}
\end{equation}
where $|\alpha|^2=(N_2)/(1+N_1)$, $|\beta|^2=(N_3)/(1+N_1)$, and
$N_1$, $N_2$ and $N_3$ are the (time dependent) average number of
quanta of the three oscillators [see Eqs.(\ref{N1}-\ref{N3})].
Since we start from vacuum we have $N_1 = N_2+N_3$ at any time.
\par
Eq. (\ref{state:0}) shows that, in general, the system is
entangled and that the distribution over the different occupation
numbers, $N_1$,$N_2$ and $N_3$, is thermal. In particular, for $
N_3\ll N_1\sim N_2$, the state $|\Psi\rangle$ reduces to
$|\psi_{12}\rangle=\frac{1}{\sqrt{1+N_1}} \sum_{n=0}^\infty
\alpha^n|n,n,0\rangle$, showing maximal entanglement between atoms
with different momenta. On the other hand, for $ N_2\ll
N_1\approx N_3$, $|\Psi\rangle$ reduces to
\begin{eqnarray}
|\psi_{13}\rangle&=&\frac{1}{\sqrt{1+N_1}} \sum_{n=0}^\infty
\beta^n|n,0,n\rangle.\label{state13}
\end{eqnarray}
showing maximal entanglement between atoms and photons.
Both the states $|\psi_{12}\rangle$ and $|\psi_{13}\rangle$
are pure bipartite states. They are maximally entangled
states for the given number of quanta, according to the
excess von Neumann entropy criterion \cite{pho}, whereas the
presence of a third mode reduces, in general, the entanglement
between the other two modes \cite{bur}.
The atom-radiation entangled state (\ref{state13}) is what
supports our teleportation scheme. Incidentally, it has the same
form of the twin-beam state of radiation used to realize continuous
variable optical teleportation \cite{kim}, and this will allows us to
employ the same kind of Bell measurement scheme.
\par
In the quantum limit $g\sqrt{N}\ll 2\omega_r$ and for $2g\sqrt N t\gg 1$
the population of the three oscillators are given by
\begin{eqnarray}
N_1(t)&\approx&\frac{1}{4}\left[1+\left(\frac{g\sqrt
N}{2\omega_r}\right)^2\right]e^{2g\sqrt N t},
\label{N1}\\
N_2(t)&\approx&\frac{1}{4}\left(\frac{g\sqrt
N}{2\omega_r}\right)^2 e^{2g\sqrt N t},
\label{N2}\\
N_3(t)&\approx&\frac{1}{4}e^{2g\sqrt N t}. \label{N3}
\end{eqnarray}
so that $N_1\approx N_3\gg N_2$. Furthermore, maximal entanglement
between modes $1$ and $3$ requires $N_2\le 1$, so that the
interaction time must satisfy the following limits
\begin{equation}
\frac{1}{g\sqrt N}\ll t_{int} \le \frac{1}{g\sqrt N}{\rm
ln}\left(\frac{4\omega_r}{g\sqrt N}\right)\:. \label{time}
\end{equation}
The state $\sigma$ we want to teleport onto the atomic mode $a_1$
pertains to an additional radiation mode $a_4$. The Bell measurement
is jointly performed on $a_3$ and $a_4$, and consists in the
measurement of the real and the imaginary part of the complex
operator $Z=a_3+a_4^\dag$. The measurement of $Z_R=\hbox{Re}[Z]$ and
$Z_I=\hbox{Im[Z]}$ corresponds to measuring the sum- and difference-quadratures
$x_3+x_4$ and $y_3-y_4$ of the two modes, where the quadrature $x$
of a mode $b$ is the operator $(b+b^{\dag})/2$, and the quadrature
$y$ is the operator $(b-b^{\dag})/2i$. Such a measurement can be
experimentally implemented by multiport homodyne detection
({\em i.e} by mixing the two modes in balanced beam splitter and
then measuring two conjugated quadratures on the outgoing modes, see
Fig. \ref{f:schema}), if the
two modes have the same frequencies \cite{msa,tri}, or by
heterodyne detection otherwise \cite{het}. The measurement is
described by the following probability operator-valued measurement
(POVM) \cite{RSP}, acting on the Hilbert space of mode $a_{3}$
\begin{equation}\label{POVM:tele}
{\Pi}_\alpha = \frac{1}{\pi}
{D}(\alpha) \:\sigma^T \: {D}^{\dag}(\alpha)
\end{equation}
where $\alpha$ is a complex number, $D(\alpha)$ is the
displacement operator $D(\alpha)=\exp\{\alpha
a_{3}^{\dag}-\bar{\alpha}a_{3}\}$
and $(\cdots)^{T}$ denotes the transposition operation. The
result of the measurement is classically transmitted to the
receiver's location (Bob), where a displacement operation
$D(\alpha)^{\dag}$ is performed on the conditional state
$\varrho_\alpha$ (see below on how to implement coherent atomic
displacement).  The dynamics of the conditional measurement is
described by  \cite{RSP}
\begin{eqnarray}
p_\alpha &=& \hbox{Tr}_{13} \left[ |\psi_{13}\rangle \langle
\psi_{13}| {\mathbb I}_{1}\otimes \Pi_{\alpha}
\right]\\
\varrho_\alpha &=& \frac1{p_\alpha} \hbox{Tr}_3
\left[|\psi_{13}\rangle \langle \psi_{13}|
{\mathbb I}_{1}\otimes \Pi_{\alpha}\right] \nonumber \\
\tau_\alpha &=& D(\alpha)\varrho_\alpha D^\dag (\alpha)\nonumber
\label{condalpha}\;,
\end{eqnarray}
where $p_\alpha$ is the probability for the result $\alpha$ in the
Bell measurement, ${\mathbb I}$ the identity operator,
$\varrho_\alpha$ is the conditioned state of the atomic beam after the
measurement, and $\tau_\alpha$ is is the conditioned state
after the displacement operation.
The teleported state is the average over all the possible
outcomes, {\em i.e.}
\begin{eqnarray}
\tau &=& \int_{\mathbb{C}} d^2\alpha \: p_\alpha \: \tau_\alpha
\label{teleported} \\ &=&
\int_{\mathbb{C}} d^2\alpha \: D(\alpha)\:\hbox{Tr}_3
\left[|\psi_{13}\rangle \langle \psi_{13}|\: \mathbf{I}_{1}\otimes
\Pi_{\alpha}\right] D^\dag (\alpha)\nonumber  \;.
\end{eqnarray}
After performing the partial trace and with some algebra
\cite{RSP}, one has
\begin{eqnarray}
\tau = \int \frac{d^2\alpha}{\pi K}\:
\exp\{-\frac{|\alpha|^2}{K}\} \: D(\alpha)\sigma D^\dag (\alpha)
\label{tel}\;,
\end{eqnarray}
where $K=1+N_{1}+N_{3}-\sqrt{(N_{1}+N_{3})(N_{1}+N_{3}+2)}$.
Eq. (\ref{tel}) shows that the overall dynamics of our scheme
is equivalent to that of a Gaussian noisy channel with parameter
$K$ \cite{ban,RSP}. The density matrix $\tau$, describing the final
state of the atomic mode $a_1$ coincides with $\sigma$ in the limit
$N_{1}+N_{3}\longrightarrow \infty$ {\em i.e.} for high gain in the
CARL dynamics. Notice that $N$, and in turn $N_1$ and $N_3$, may vary 
in the repeated preparations of the condensate, thus introducing 
additional noise in the teleported state. 
\par
The displacement operation $D(\alpha)$ that should be performed on
the conditional atomic state $\varrho_\alpha$ can be obtained using
the same CARL Hamiltonian in the condensate, by
injecting a suitably modulated pulse, {\em i.e.} by exciting the mode
$a_3$ in a classical coherent state. In this case, assuming a short
pulse, the effective Hamiltonian may be written as $H_2 = i g \sqrt{N}
\gamma (a_1 + a_2 + h.c)$ where $\gamma$ is the amplitude of the modulated pulse.
The terms proportional to $a^\dag_j a_j$, $j=1,2$ in (\ref{ham3}) can be
discarded, and the evolution operator $U=\exp(iH_2\tau)=D_1(\alpha) \otimes
D^\dag_2(\alpha)$ coincides with the product of two displacement operators,
one for each of the atomic modes, with amplitude given
by $\alpha=-g\bar\gamma\sqrt{N}\tau$, where $\tau$ is an effective interaction
time. The amplitude $\gamma$ of the pulse can suitably tuned to obtain the
desired value of the amplitude $\alpha$, matching the results of the Bell
measurements. The above dynamics displaces both the atomic modes, however
without introducing quantum correlations. Therefore, we just ignore the effect
on the atomic mode $a_2$, which does not participate to the teleportation
protocol.
\par
The time duration of the pulse should be small when compared to
the time scale of the CARL dynamics and the decoherence time of
$|\psi_{13}\rangle$ under free evolution. This is in order for two
reasons: on one hand we have that the CARL dynamics should be switched
off after producing the desired entangled state $|\psi_{13}\rangle$, and
therefore the whole protocol should be completed within the
decoherence time. On the other hand, the displacement should be performed
on a time scale comparable to that of the Bell measurement, {\em i.e} 
before the {\em reset} of the dynamics and the generation of the subsequent 
copy of the atom-radiation entangled state in the new condensate by CARL. 
Overall, our protocol may be described as a feed forward control scheme, 
randomized according to the statistics of the Bell measurement.
\par
In order to discuss the readout part of our scheme, we should go
back to the initial entangled scheme produced by the CARL
dynamics. This should be more properly written as
$|\psi_{13}\rangle = (1+N_1)^{-1/2}\sum_{n=0}^N \beta^n
|n,0,n,N-n>$ where the fourth entry in the ket describes the
number of atoms in the condensate. Since N is a large number (of
the order of $10^7$) writing the state as in Eq.
(\ref{state:0}) is perfectly admissible as far as we are concerned
with its dynamics. However, this should be taken into account if
we want to reconstruct the state of the output atomic beam. Let us
consider, for the sake of simplicity, the initial light signal
state as a pure state $\sigma=|\varphi\rangle\langle\varphi|$,
$|\varphi\rangle=\sum_n \varphi_n\:|n\rangle$. In the limit of
high CARL gain the teleported state on the atomic beam is given by
$|\varphi^\prime\rangle=\sum_n \varphi_n |n,N-n\rangle$. This
indicates that any proper verification of the teleportation should
involve a measurement also on the condensate, {\em e.g.} a two
mode tomographic method involving both the measurement of both
momentum-mode and condensate quadratures \cite{rev}. Such kind of
measurements are at present experimentally challenging and
therefore, in order to obtain an experimentally accessible readout
system, we propose to check only the statistics of the population
$|\varphi_n|^2$, {\em i.e.} the diagonal part of the teleported
state, which can be achieved by counting atoms. If we choose
$|\varphi\rangle$ as a squeezed vacuum or a Fock number state, we
end up with an even-odd statistics or with a very sharp
atomic number distribution. This should be convincing enough, 
in order to show that nonlocal correlations between the input 
radiation mode and the output atomic mode has been established and 
exploited.
\par
In conclusion, we have proposed a novel scheme to realize the interspecies
teleportation of the quantum state of a single mode radiation field onto the
collective state of atoms with a given momentum out of a Bose-Einstein condensate.
The entangled resource needed for the teleportation protocol is established
through the CARL interaction of a single mode with the condensate in presence of a
strong far off-resonant pump laser, whereas the coherent atomic displacement is
obtained through the same interaction by injecting a suitably modulated short
coherent pulse. In principle, verification of the teleportation scheme requires
a joint measurement on the recoiling atoms and the condensate. However, for
diagonal matrix elements in the photon representation only atom counting
measurements are needed.
\par
This work has been sponsored by INFM through the project
PRA-2002-CLON and by MIUR through the PRIN projects {\em
Entanglement assisted high precision measurements} and {\em
Coherent interaction between radiation fields and Bose-Einstein
condensates}. MGAP is research fellow at {\em Collegio Alessandro
Volta}. We thank B. Englert for stimulating discussions.
%%%%%%%%%%%%%%%%%%%%%%%%%%%%%%%%%%%%%%%%%%%%%%%%%%%%%%%555

%%%%%%%%%%%%%%%%%%%%%%%%%%%%%%%%%%%%%%%%%%%%%%%%%%%%%%%555
\end{document}